\documentclass[12pt]{iopart}
\usepackage{iopams,verbatim}

\def\Journal#1#2#3#4{{#4} {\it #1} {\bf #2}, #3 }
\def\a{\alpha}
\def\b{\beta}

\def\d{\delta}
\def\e{\epsilon}
\def\diver{\textrm{div}}
\def\curl{\textrm{curl}}

\begin{document}

\title{Purely radiative perfect fluids}

\author{B Bastiaensen, H R Karimian, N Van den Bergh and L~Wylleman}

\address{Faculty of Applied Sciences TW16, Gent University, Galglaan 2, 9000 Gent, Belgium}

\begin{abstract}
We study `purely radiative' (i.e.~$\diver E =\diver H=0$) and
geodesic perfect fluids with non-constant pressure and show that the
Bianchi class A perfect fluids can be uniquely characterized
---modulo the class of purely electric and (pseudo-)spherically
symmetric universes--- as those models for which the magnetic and
electric part of the Weyl tensor and the shear are simultaneously
diagonalizable. For the case of constant pressure the same
conclusion holds provided one also assumes that the fluid is
irrotational.
\end{abstract}

\pacs{04.20.Jb, 04.40.Nr}

\section{Introduction}

 In a previous paper~\cite{dust} we discussed `purely
radiative' (i.e.~$\diver E =\diver H=0$) irrotational dust
spacetimes for which the magnetic part of the Weyl tensor is
diagonal in the shear-electric eigenframe. We show that this
analysis can be generalized to `purely radiative' and geodesic
perfect fluids. These are solutions of the Einstein field equations
\begin{equation}
    R_{ab}-\frac{1}{2}R g_{ab} = (\rho + p)u_a u_b+ pg_{ab} \ \ (\rho+p\neq0), %\quad u_au^a=-1
\end{equation}
where $u^a$ is the normalized $4$-velocity, $\rho$ and $p$ the
density and pressure of the fluid and $g_{ab}$ the spacetime metric,
and for which we impose the additional restrictions (1) that the
flow is geodesic: $\dot{u}_a = u_{a;b}u^b=0$ and (2) that the
electric and magnetic parts $E_{ab}$ and $H_{ab}$ of the Weyl tensor
have vanishing \emph{spatial} divergence. The spatial tensors
$E_{ab}$ and $H_{ab}$ are hereby defined by
\begin{equation}\label{EenH}
E_{ab}\equiv C_{acbd}u^c u^d=E_{<ab>}   \quad  H_{ab}\equiv \frac{1}{2}{\varepsilon_{acd}}{C^{cd}}_{be}u^e=H_{<ab>},
\end{equation}
where $\varepsilon_{abc}=\eta_{abcd}u^d$ is the spatial projection
of the spacetime permutation tensor $\eta_{abcd}$ and
$S_{<ab>}={h_a}^c{h_b}^dS_{(cd)}-\frac{1}{3}S_{cd}h^{cd}h_{ab}$ is
the projected, symmetric and trace-free part of $S_{ab}$, with
$h_{ab}=g_{ab}+u_au_b$ the spatial projector into the comoving rest
space. The covariant spatial divergence and curl for tensors are
defined by \cite{Maartens2}:
\begin{equation}\label{divcurl}(\diver S)_a = D^b S_{ab}, \quad \curl S_{ab}=\varepsilon_{cd(a}D^c{S_{b)}}^d,
\end{equation}
where the spatial part $D_a$ of the covariant derivative $\nabla_a$ is given by
\[D_aA_{b\cdots}={h_a}^c{h_b}^d\cdots\nabla_cA_{d\cdots}.
\]
Gravitational radiation is covariantly described by the nonlocal
fields $E_{ab}$ (the tidal part of the curvature), which generalizes
the Newtonian tidal tensor, and by $H_{ab}$, which has no Newtonian
analogue~\cite{Ellis}. As such, $H_{ab}$ may be considered as the
true gravitational wave tensor, since there is no gravitational
radiation in Newtonian theory. However at least in the linear
regime, as in electromagnetic theory, gravity waves are
characterized by $H_{ab}$ and $E_{ab}$, where both are
divergence-free but neither is curl-free~\cite{Hogan}. For this
reason spacetimes in which $\diver E=0=\diver H$ have been termed
\emph{purely radiative}~\cite{Sopuerta}.

Although, when considering the gravitational dynamics of matter in
the full non-linear regime, these conditions give rise to two
independent chains of severe integrability conditions, it was
remarked in \cite{Sopuerta} that both are satisfied by the Bianchi
class A models. This motivated us to investigate purely radiative
perfect fluids in their full generality.

As a first step we have shown that in fact all the non-rotating dust
models belong to Bianchi class A, provided we assume that they are
not purely electric (in which case the term `purely radiative' would
be inappropriate anyway). In the purely electric case the exceptions
were given by the Petrov type D spatially homogeneous and locally
rotationally symmetric (LRS) metrics which are pseudo-spherically
symmetric (and hence of Bianchi type $III$), or which are
spherically symmetric and belong to the Kantowski-Sachs family of
dust models and hence do not admit a 3 dimensional isometry group
acting simply transitively on the hypersurfaces of homogeneity. For
technical reasons we also assumed that $[\sigma,\, H] = 0$. Together
with the fact that $[\sigma,\, E] = 0$, which is a consequence of
$\diver H = 0$, this implies that $\sigma_{ab}$, $E_{ab}$ and
$H_{ab}$ can be simultaneously diagonalized (it is not necessary to
assume that $[E,\, H]=0$, as a degenerate shear eigenplane also must
be a degenerate $E$ eigenplane, see section 5).

When, on the contrary, one considers fluids with non-constant
pressure (but still with geodesic flow), the spatial gradient of $p$
vanishes and the velocity becomes orthogonal to the $p=$ constant
hypersurfaces, such that the flow is
irrotational~\cite{RaychMaity,Synge}. It is then a straightforward
consequence of the assumptions $\diver E = 0$ and $[\sigma,\, H] =
0$ that the spatial gradient of the matter density $\rho$ vanishes
too. As stated in \cite{RaychMaity} it follows that $p$, $\sigma$
and $\theta$ are constants over the hypersurfaces of constant
density and hence the resulting spacetimes satisfy the so called
Postulate of Uniform Thermal Histories~\cite{Bonnor, BonnorPugh,
Barry, Norbert}. This seems to \emph{suggest} that a local group of
isometries exists, mapping the flow lines into each other (so called
observational homogeneity). However not much progress has been made
in this area and a detailed proof, or even a precise formulation of
the conjecture that PUTH would lead to observational homogeneity, is
still lacking. In fact it was suggested in \cite{BonnorPugh} that
any extra distinguishing mathematical property leading to
observational homogeneity probably would have to be very
complicated, involving at least third-order derivatives of the
metric. An example of such an extra property is precisely the
condition of being purely radiative, in the above sense of having
$\diver E=0=\diver H$, which indeed enables us to show that the
resulting spacetimes are spatially homogeneous and, more
particularly, of Bianchi class A when we exclude the `non-radiative'
purely electric subcases of section \ref{sec:LRS}.

\section{Basic variables and equations}\label{sec:basic}

We follow the notations and conventions of the Ellis-MacCallum
orthonormal tetrad formalism~\cite{MacCallum}. Herein, the
normalized 4-velocity $u^a$ plays the role of the timelike basis
vectorfield of the tetrad $\{\mathbf{e}_0,\mathbf{e}_\alpha\}$.
Greek indices take the values 1, 2 and 3 and refer to tetrad
components with respect to $\mathbf{e}_{\alpha}$.

The basic variables in the formalism are 24 independent linear
combinations of the Ricci rotation coefficients or, equivalently, of
the commutator coefficients $\gamma^a_{bd}$ associated with the
tetrad; these are the objects $n_{\a\b}$ and $a_\a$ defined by
$\gamma^\a_{\b\d}=\epsilon_{\b\d\e}n^{\e\a}+\delta^\a_\d a_\b -
\delta^\a_\b a_\d$, the components $\Omega_\a$ of the angular
velocity of the triad $\mathbf{e}_\a$ with respect to the local
`inertial compass' and the components $\dot{u}_\a$,
$\omega_{\alpha}$ and $\theta_{\a\b}$ of the spatial kinematic
quantities (acceleration vector, vorticity vector and expansion
tensor respectively). We also use the decomposition
$\theta_{\a\b}=\sigma_{\a\b}+\theta/3 \delta_{\a\b}$, with
$\sigma_{\a\b}$ the trace-free shear tensor and $\theta$ the expansion
scalar. It turns out that there is a slight advantage in
using instead of the variables $n_{\alpha \beta} $ and $a_\alpha$
the connection coefficients $n_\alpha$, $q_\alpha$ and $r_\alpha$
related to the former by
\begin{equation*}
n_{\alpha +1\ \alpha-1}=(r_\alpha+q_\alpha)/2,\quad
a_\alpha=(r_\alpha-q_\alpha)/2,\quad n_{\alpha
\alpha}=n_{\alpha+1}+n_{\alpha-1}
\end{equation*}
(these expressions have to be read modulo 3, so for example $\alpha
= 3$ gives $n_{12}= q_3 + r_3$). For sake of simplicity we define
$Z_{\alpha}$, $m_\alpha$ and $S_\alpha$ as the spatial gradients of
$\theta$, $\rho$ and $\sigma^2\equiv \sigma^{ab}\sigma_{ab}$
respectively: $Z_{\alpha}= \partial_{\alpha}\theta$,
$m_\alpha=\partial_{\alpha}\rho$ and
$S_\alpha=\partial_\alpha\sigma^2$.

Together with the Einstein field equations, the basic equations are
the Bianchi equations and the Ricci equations (part of which is
equivalent with the Jacobi equations), which we have included in the
appendix.

\section{Generalities}\label{sec:gen}
From the Bianchi identity (\ref{Bianchi3}) for geodesic perfect
fluids (see appendix) and the assumption $\diver H=0$ we get,
\[
 (\diver{H})_\alpha\equiv D^\beta H_{\alpha\beta}=0\Longleftrightarrow[{E},\sigma]=0,
\]
which guarantees the existence of a frame wherein $E_{\alpha\beta}$ and $\sigma_{\alpha\beta}$ are both diagonal and which will be used henceforth.

Because all off-diagonal elements of $E_{\alpha\beta}$ vanish we derive from (\ref{Omegas}) that:
\begin{equation}\label{angular}\left.\begin{array}{lll}
\Omega_1(\sigma_{22}-\sigma_{33})~=\Omega_2(\sigma_{33}-\sigma_{11})~=\Omega_3(\sigma_{11}-\sigma_{22})=0.\\
\end{array}\right.
\end{equation}
As a consequence the angular velocity $\Omega_{\alpha}$ can be assumed
to be 0: either all components automatically vanish, or, when e.g.~the $(e_2,e_3)$ plane is a shear
eigenplane, then $\Omega_2=\Omega_3=0$ and we can choose an extra
rotation in this plane to make $\Omega_1=0$. The latter operation leaves the frame then fixed
up to rotations in the $(e_2,e_3)$-plane for which the rotation
angle $\varphi$ satisfies $\partial_0\varphi=0$. In fact, under a rotation about an angle $\varphi$ in the $(e_2,e_3)$-plane the quantities $\Omega$ transform as follows:
\[\Omega^{'}_1,\Omega^{'}_2,\Omega^{'}_3 \longrightarrow \Omega_1-\partial_0\varphi, \Omega_2\cos \varphi-\Omega_3\sin \varphi, \Omega_2\sin \varphi+\Omega_3\cos \varphi.
\]

The Raychaudhuri equation (\ref{Ein00}) and the conservation law
(\ref{Bianchi1}) give us the evolution of respectively the expansion
and the density,
%: \begin{equation}\label{dotThetaDotRho}  \begin{array}{lll}\partial_0\theta&=&-\frac{1}{3}\theta^2-\sigma^2-\frac{1}{2}(\rho+3p)\\
%\partial_0\rho&=& -(\rho+p)\theta,
%\end{array}
%\end{equation}
while (\ref{orthdefE1}) gives
the evolution of the components $\sigma_{\alpha\alpha}$ :
\begin{equation}\label{dotSigma}
\eqalign{
3\partial_0\sigma_{11}= (\sigma_{22}^2+\sigma_{33}^2-2\sigma_{11}^2-{2}\sigma_{11}\theta)-3E_{11},\\
3\partial_0\sigma_{22}= (\sigma_{33}^2+\sigma_{11}^2-2\sigma_{22}^2-{2}\sigma_{22}\theta)-3E_{22}, \\
3\partial_0\sigma_{33}= (\sigma_{11}^2+\sigma_{22}^2-2\sigma_{33}^2-2\sigma_{33}\theta)-3E_{33}.
}
\end{equation}

The spatial derivatives of $\sigma_{\alpha\alpha}$ are obtained by the $(0\alpha)$-Einstein field equations (\ref{Ein0a}) and by the off-diagonal components of (\ref{Ricci1}) :
\begin{equation}\label{dsigma} \left. \begin{array}{lll}
\partial_1\sigma_{11}&=&\frac{2}{3}Z_1+r_1(\sigma_{11}-\sigma_{22})-q_1(\sigma_{11}-\sigma_{33}),\\
%\partial_2\sigma_{22}&=&\frac{2}{3}Z_2+r_2(\sigma_{22}-\sigma_{33})-q_2(\sigma_{22}-\sigma_{11}),\\
%\partial_3\sigma_{33}&=&\frac{2}{3}Z_3+r_3(\sigma_{33}-\sigma_{11})-q_3(\sigma_{33}-\sigma_{22}),\\
%\partial_1\sigma_{22} &=& -\frac{1}{3}Z_1+r_1(\sigma_{22}-\sigma_{11})+H_{23},\\
%\partial_1\sigma_{33} &=&-\frac{1}{3}Z_1+q_1(\sigma_{11}-\sigma_{33})-H_{23},\\
\partial_2\sigma_{11} &=& -\frac{1}{3}Z_2+q_2(\sigma_{22}-\sigma_{11})-H_{13},\\
%\partial_2\sigma_{33} &=& -\frac{1}{3}Z_2+r_2(\sigma_{33}-\sigma_{22})+H_{13},\\
\partial_3\sigma_{11} &=& -\frac{1}{3}Z_3+r_3(\sigma_{11}-\sigma_{33})+H_{12}\\
%\partial_3\sigma_{22} &=& -\frac{1}{3}Z_3+q_3(\sigma_{33}-\sigma_{22})-H_{12}\\
\end{array} \right.
\end{equation}

and similar expressions by cyclic permutation of the indices. Note
that the $\diver{E}=0$ condition relates the off-diagonal parts of
$H_{\alpha\beta}$ to $m_\alpha$ by (\ref{Bianchi2}):
\begin{equation}\label{Bianchi22} \left. \begin{array}{lll}
m_1&=&3H_{23}(\sigma_{33}-\sigma_{22}),\\
m_2&=&3H_{13}(\sigma_{11}-\sigma_{33}),\\
m_3&=&3H_{12}(\sigma_{22}-\sigma_{11}).\\
\end{array} \right.
\end{equation}
Using (\ref{Bianchi1}) we can now act with the commutator
$[\partial_0,\partial_{\alpha}]\equiv
-(\sigma_{\alpha\alpha}+\frac{1}{3}\theta)\partial_\alpha$ on $\rho$
to obtain the relation:
\begin{equation}\label{dotma}
\partial_0 m_\alpha = -m_\alpha(\sigma_{\alpha \alpha}+\frac{4}{3} \theta) -(\rho+p) Z_\alpha, \quad \alpha=(1,2,3).
\end{equation}

Propagating (\ref{Bianchi22}) along the fluid flow and substituting (\ref{dotSigma}) and (\ref{dotma}) gives then the following expressions:
\begin{equation}\label{hdot}
\begin{array}{lll}
\fl3(\sigma_{11}-\sigma_{22})\partial_0 H_{12} = (6\sigma_{11}^2-6\sigma_{22}^2-6E_{22}-3E_{33}+2\sigma_{22}\theta-2\sigma_{11}\theta)H_{12}+(\rho+p)Z_3\\

\fl3(\sigma_{22}-\sigma_{33})\partial_0 H_{23} = (6\sigma_{22}^2-6\sigma_{33}^2-6E_{33}-3E_{11}+2\sigma_{33}\theta-2\sigma_{22}\theta)H_{23}+(\rho+p)Z_1\\

\fl3(\sigma_{33}-\sigma_{11})\partial_0 H_{13} =
(6\sigma_{33}^2-6\sigma_{11}^2-  6
E_{11}-3E_{22}+2\sigma_{11}\theta-2\sigma_{33}\theta)H_{13}+(\rho+p)Z_2.
\end{array}
\end{equation}

Combining the equations above with the spatial information in the
Bianchi and Jacobi equations allows one to obtain \emph{explicit}
algebraic expressions for \emph{all} the directional derivatives of
the variables $p, \rho, \theta, \sigma_{\a\a}, Z_\a, E_{\a\a},
H_{\a\b},r_\a,q_\a$ and $n_\a$. Although the integrability
conditions for the resulting Pfaffian system are polynomial and
although the complicated nature of these restrictions seems to
suggest that solutions can only be spatially homogeneous, the size
of the polynomials involved and the large number of different
subcases left to be investigated, make it impossible so far to draw
any general conclusions. Therefore we will restrict in the present
paper to one of the main subcases, namely $[\sigma,\, H]=0$. As
mentioned in the introduction this assumption implies that
$H_{\alpha\beta}$ is diagonal in the $(\sigma, E)$-eigenframe (in
the case of degenerate shear (\ref{orthdefE1}) implies that the
eigenplanes of $\sigma_{\alpha \beta}$ and $E_{\alpha \beta}$
coincide). From (\ref{Bianchi22}) and (\ref{dotma}) or (\ref{hdot})
one obtains then $m_{\alpha}=0$ and $Z_\alpha=0$ (regardless of the
degeneracy of $\sigma_{\alpha\beta}$), giving a first hint that the
corresponding spacetimes might be spatially homogeneous indeed. To
prove this we first construct some information about the remaining
part of the magnetic tensor.

Using the $[\partial_0,\partial_{\alpha}]$ commutators on $\theta$,
we obtain from (\ref{Ricci3}) and $Z_\alpha=m_\alpha=\partial_\alpha
p=0$ that $S_\alpha\equiv\partial_\alpha \sigma_{\beta
\gamma}\sigma^{\beta \gamma} =0$ and hence, using (\ref{dsigma}),

\begin{equation}\label{rq1}
\begin{array}{lll}
r_1(\sigma_{11}-\sigma_{22})^2-q_1(\sigma_{11}-\sigma_{33})^2 &=& 0,\\
r_2(\sigma_{22}-\sigma_{33})^2-q_2(\sigma_{22}-\sigma_{11})^2 &=& 0,\\
r_3(\sigma_{11}-\sigma_{33})^2-q_3(\sigma_{33}-\sigma_{22})^2 &=& 0 .\\
\end{array}
\end{equation}

We also need the spatial derivatives of $E_{\alpha\alpha}$, which we
can deduce from (\ref{Bianchi2}) and the off diagonal part of
(\ref{Bianchi5}) :
\begin{equation}\label{defdiv} \left. \begin{array}{lll}
\partial_1E_{11}&=& E_{11}(r_1-2q_1)-E_{22}(r_1+q_1), \\
%\partial_2E_{22}&=& E_{22}(r_2-2q_2)-E_{33}(r_2+q_2), \\
%\partial_3E_{33}&=& E_{33}(r_3-2q_3)-E_{11}(r_3+q_3), \\
\partial_2 E_{11} &=& (E_{22}-E_{11})q_2,\\
\partial_3 E_{11} &=& (E_{11}-E_{33}) r_3.
\end{array} \right.
\end{equation}

We will treat the two cases of non-degenerate and degenerate shear
now separately. Note that by (\ref{dotSigma}) and (\ref{Ricci21})
$\sigma_{\alpha \beta}\neq 0$, as otherwise the solutions would be
FLRW.

\section{Non degenerate shear}
As the frame is invariantly defined, we first aim to proof that $r_\alpha$ and $q_\alpha$ are zero and
that all spatial derivatives vanish. We first look at the evolution of $\sigma^2$: from (\ref{Ricci3}) it is easy to check that
\begin{equation}\label{dotSigma2}
\partial_0\sigma^2 =-\frac{4}{3}\theta\sigma^2- 2 \sigma^3+ 2 E\cdot\sigma,
\end{equation}
where $\sigma^3\equiv \sigma_{\gamma \alpha}{\sigma_\beta}^{\gamma}\sigma^{\alpha\beta}$ and $E\cdot\sigma\equiv E_{\alpha\beta}\sigma^{\alpha\beta}$.

Hence, acting with the commutator $[\partial_0,\partial_\alpha]$ on
$\sigma^2$ we find:
\begin{equation}\label{dotS2}
\partial_0S_\alpha = \frac{5}{3}\theta S_\alpha-\frac{4}{3}\sigma^2 Z_\alpha-2\partial_\alpha\sigma^3-2\partial_\alpha (E\cdot\sigma)-S_\alpha\sigma_{\alpha\alpha}
\end{equation}
and hence, as $Z_\alpha=S_\alpha=0$, $\partial_\alpha\sigma^3+\partial_\alpha (E\cdot\sigma)=0$. Working out the latter componentwise and subsituting
%\[
%\sum_{\delta=1}^3(\sigma_{\delta\delta}\partial_\alpha E_{\delta\delta}+
%E_{\delta\delta}\partial_\alpha\sigma_{\delta\delta}+
%3\sigma_{\delta\delta}^2\partial_\alpha\sigma_{\delta\delta})=0.\]
(\ref{defdiv}) and (\ref{dsigma}), results in three algebraic expressions. For $\alpha=1$ we get
\begin{equation}\label{relrq}
\fl r_1(\sigma_{22}-\sigma_{11})[4E_{22}+2E_{33}-3\sigma_{33}(\sigma_{22}-\sigma_{11})]-
q_1(\sigma_{33}-\sigma_{11})[2E_{22}+4E_{33}-3\sigma_{22}(\sigma_{33}-\sigma_{11})]=0.
\end{equation}
Eliminating $r_1$ from (\ref{rq1}) and (\ref{relrq}), we find
\begin{equation}\label{voorwQ}q_1(\sigma_{33}-\sigma_{11})(\sigma_{22}-\sigma_{11})\chi=0
\end{equation}with
\[\fl\chi\equiv\frac{2}{3}[E_{11}(\sigma_{22}-\sigma_{33})+E_{22}(\sigma_{33}-\sigma_{11})+E_{33}(\sigma_{11}-\sigma_{22})]
-(\sigma_{11}-\sigma_{22})(\sigma_{22}-\sigma_{33})(\sigma_{33}-\sigma_{11}).
\]
The whole calculation can be repeated for the cases $\alpha=2,3$ so that we finally have (as the shear is non-degenerate)
\begin{equation}\label{rq2}
q_1\chi~=q_2\chi~=q_3\chi=0.\\
\end{equation}

If $\chi \neq 0$ this implies with (\ref{rq1}) that
$r_{\alpha}=q_{\alpha}=0$, while when $\chi=0$ one has $\partial_1
\chi =0$ and hence, using (\ref{dsigma}) and (\ref{defdiv}),
% If $q_1^2+q_2^2+q_3^2\neq 0$ it would follow from (\ref{rq2}) that
%$\chi=0$, after which $\partial_1\chi=0$ leads to
\begin{equation}\label{relrq2}q_1(\sigma_{33}-\sigma_{11})^3+r_1(\sigma_{22}-\sigma_{11})^3=0.
\end{equation}
%using (\ref{defdiv}) and (\ref{dsigma}).
Eliminating $q_1$ from (\ref{relrq2}) and (\ref{rq1}.a) and using
$\textrm{tr}(\sigma_{\alpha \beta})=0$
%gives \[-3r_1(\sigma_{22}+\sigma_{33})(\sigma_{33}-\sigma_{11})^2(\sigma_{22}-\sigma_{11})^2=0,
%\]
%which, as (\ref{rq1}) implies that $r_1^2+r_2^2+r_3^2\neq 0$, leads to $\sigma_{11}=
%-(\sigma_{22}+\sigma_{33})=0$. Similarly one would obtain $\sigma_{22}=\sigma_{33}=0$ and hence $\sigma_{\alpha \beta}=0$.
again leads to $r_{\alpha}=q_{\alpha}=0$.
%We conclude that for non-degenerate shear $q_{\alpha}=0$ and hence,
%by (\ref{rq1}), also $r_{\alpha}=0$.

This means that $a_{\alpha}=0$ and $n_{\alpha\beta}$ is diagonal. It
is easy to check that all spatial derivatives of the rotation
coefficients now vanish and hence~\cite{Ellis3} we obtain the
non-degenerate spatially homogenous Bianchi class A models,
i.e.~types $I, II,VI_0,VII_0,VIII$ and $IX$.

\section{Degenerate shear}
From the degeneracy of the shear it is easily verified that $E_{\alpha\beta}$ must be degenerate too (\ref{orthdefE1}) and we may assume
\[\begin{array}{lll}\sigma_{\alpha\beta}&=&\textrm{diag}(-2\sigma,\sigma,\sigma)\\
  E_{\alpha\beta}&=&\textrm{diag}(-2E,E,E).
  \end{array}
\]

For dust we can discard the case of vanishing $E$ as a consequence
of the non-existence of anti-Newtonian universes~\cite{Lode}, but
for non-constant pressure purely magnetic perfect fluids, which are
necessarily of Petrov type I or D,  do exist (we don't consider the
conformally flat cases, as these uniquely reduce under the present
assumptions to the FLRW models, see \cite{Kramer}): for type I the
only possibility is the Bianchi $VI_0$ metric discussed in
\cite{WyVdBPRD}, while for type D the spacetime is locally
rotationally symmetric class III~\cite{WyVdBPRD}. As the fluid is
non-rotating the solutions are then given precisely by the Bianchi
type $VIII$ or $IX$ Lozanovski-Aarons-Carminati (LAC)
metrics~\cite{Lozanovski, Lozanovski2,LozanovskiAa}, with a possible
degeneracy to Bianchi type II (see below).
%a and contains as a special case the $p = \rho/5$
%Collins-Stewart metric~\cite{CollinsStewart}.

Henceforth we take $E$ non zero (both the type I and type D purely
magnetic models above will reappear as $E=0$ special cases).
Assuming that $H_{\alpha\beta}$ is diagonal in the
$(\sigma,E)$-eigenframe, $H_{\alpha
\beta}=\textrm{diag}(-H_{22}-H_{33},H_{22},H_{33})$, (\ref{rq1})
implies

\begin{equation}\label{Hcond}
r_1-q_1 = r_3 = q_2 = 0,
\end{equation}
while the (23)-component of (\ref{Ricci2}) shows that $r_1+q_1=0$
and hence $r_1=q_1=0$.

At this point the only variables having possibly non-vanishing
spatial derivatives are $q_3,r_2$ and $n_1,n_2,n_3$.
With the simplifications obtained so far, the ($11$)-Einstein equation (\ref{Einaa}) reduces to
\begin{equation}\label{defK} 0=\frac{1}{3}\rho+\frac{1}{3}\theta(\sigma-\frac{1}{3}\theta)+\frac{2}{3}\sigma^2-E-n_2 n_3.
\end{equation}
Eliminating now
$\dot E_{\alpha \beta}$ from (\ref{Bianchi4}) for $\alpha=\beta=1$
and $\alpha=\beta=2$, we obtain
\[0=H_{11}(-2n_1+n_2-2n_3)+H_{33}(-4n_1-n_2-n_3),
\]which in combination with (\ref{Ricci21}) leads to
\begin{equation}\label{vwn}(n_{1}+n_{2}+n_{3})(n_{2}-n_{3})=0.
\end{equation}
We will discuss the three cases that follow from (\ref{vwn})
separately; i.e.~$n_2\neq n_3$, $n_2=n_3\neq 0$ and $n_2=n_3=0$. Notice that when $n_2=n_3$ the necessary and sufficient conditions for the fluid to be locally rotationally symmetric are
automatically satisfied~\cite{Goode}.

\subsection{$n_2-n_3\neq 0$}
By (\ref{vwn}) we have $n_1+n_2+n_3=0$.

From (\ref{Jac111213}) we get the evolution of $n_\alpha$, namely
\begin{equation}\label{dotN}
\begin{array}{lll}
\partial_0n_1 &=& \sigma(n_2+n_3)-\frac{1}{3}n_1\theta\\
\partial_0n_2 &=& -\sigma(n_2+3n_3)-\frac{1}{3}n_2\theta .
%\partial_0n_3 &=& -\sigma(3n_2+n_3)-\frac{1}{3}n_3\theta .
\end{array}
\end{equation}
Herewith we obtain, propagating $n_1+n_2+n_3$ along $\mathbf{e}_0$,
\[0=\partial_0(n_1+n_2+n_3) = -3(n_2+n_3)\sigma\]and thus $n_1=0$ and $n_2=-n_3\neq0$.
Applying $\partial_3$ to (\ref{defK}) we find $\partial_3n_2=0$. On
the other hand, calculating $\partial_3n_2$ from the ($12$)-Einstein
equation (\ref{Einab}) and the Jacobi equation (\ref{Jac234}),
results in $\partial_3n_2=q_3(n_3-n_2)$. Hence we see that
\[q_3(n_3-n_2)=0
\] and thus $q_3=0$, as $n_3\neq n_2$ by assumption. It follows now that also $r_2=0$ as $\partial_1q_3=r_2n_2$ by (\ref{Einab}) and
(\ref{Jac234}). So $r_{\alpha}$ and $q_{\alpha}$ all become zero and we have
\[\begin{array}{lll}
n_{\alpha\beta} &=& \textrm{diag}(0,-n_2,n_2) \quad (n_2\neq0)\\
a_{\alpha} &=& 0.
\end{array}
\]This implies the vanishing of the spatial derivatives of all rotation
coefficients ($\partial_{\alpha}\equiv0$) and we obtain a spatially
homogeneous universe of Bianchi class A, type $VI_0$. Examples are
the metrics without rotational symmetry in \cite{CollinsGlass}.

\subsection{$n_2 = n_3\neq0$}
From (\ref{defK}) and the ($33$)-Einstein equation (\ref{Einaa}) we can calculate
\begin{equation}\label{E:e2r2} \partial_2r_2\equiv \rho+3\sigma^2-\frac{1}{3}\theta^2-n_2^2-2n_2n_1+q_3^2+r_2^2+\partial_3q_3.\end{equation}
At this stage $n_{\alpha\beta}$ and $a_\alpha$ are given by
\[n_{\alpha\beta}=\left[\begin{array}{ccc}2n_2 &q_3/2&r_2/2\\
q_3/2&n_1+n_2&0\\
r_2/2&0&n_1+n_2\end{array}\right], \quad a_{\alpha} =
\left[\begin{array}{c}0\\ r_2/2\\-q_3/2\end{array}\right].\] Now
$\partial_1 n_2=0$ by the Jacobi equation (\ref{Jac234}), while
$\partial_2n_2=0$ and $\partial_3n_2=0$ follow by combining
respectively the $(31)$-Einstein equation (\ref{Einab}) and the
$(12)$-Einstein equation (\ref{Einab}) with the Jacobi equation
(\ref{Jac234}), so the spatial derivatives of $n_2$ vanish. We now
try to obtain the Bianchi A condition $a_\alpha=0$: remembering that
we still have a rotational degree of freedom left (namely rotations
in the (23)-plane about an angle $\varphi$ satisfying $\partial_0
\varphi = 0$), we use the following transformation formulas for the
quantities $n_{\alpha\beta}$ and $a_\alpha$:
$n_{\alpha\beta}\rightarrow n^r_{\alpha\beta}$, $a_\alpha
\rightarrow a^r_{\alpha}$ with

\begin{equation}\label{rotN}
\eqalign{
n^r_{11} &= n_{11}, \nonumber \\
n^r_{22}&= n^r_{33} = n_{22}-\partial_1\varphi, \nonumber \\
n^r_{12} &= \frac{1}{2}[\cos(\varphi)(\partial_2\varphi+q_3)-\sin(\varphi)(\partial_3\varphi+ r_2)], \nonumber \\
n^r_{13} &=
\frac{1}{2}[\cos(\varphi)(\partial_3\varphi+r_2)+\sin(\varphi)(\partial_2\varphi+
q_3)] }
\end{equation}

and
\begin{equation}\label{rotA}
n^r_{23}=a^r_1=a^r_2-n^r_{13}=a^r_3+n^r_{12}=0.
\end{equation}

A rotation making $a_{\alpha}=0$ and $n_{\alpha \beta}$ diagonal can
then be obtained by $\partial_2 \varphi = -q_3$ and $
\partial_3 \varphi = -r_2$, the integrability condition of which
leads to $\partial_1\varphi = -\frac{K}{2n_2}+n_1+n_2$, with
\begin{equation}\label{K}K\equiv\rho+3\sigma^2-\frac{1}{3}\theta^2+n_2^2.\end{equation}
One can check that the integrability conditions for the resulting
system of pde's
\begin{equation}\label{rot2}
\eqalign{
\partial_0\varphi &= 0, \nonumber \\
\partial_1\varphi &= -\frac{K}{2n_2}+n_1+n_2,\nonumber \\
\partial_2\varphi &= -q_3,\\
\partial_3\varphi &= -r_2 \nonumber
}
\end{equation}
are identically satisfied. For the rotated variable $n_1$ we find
$n_1+n_2= \frac{K}{2n_2}$: all spatial derivatives now vanish and we
obtain the spatially homogeneous LRS Bianchi class A spacetimes of
types $II$ ($K=0$), $VIII$ ($K<0)$ and $IX$ ($K>0$). An integrable
subcase arises when $\rho= 3 n_2^2-\theta(\sigma -\theta/3)-6 \sigma ^2$: % wel degelijk correct
the electric part of the Weyl tensor is then $0$ and we obtain the
purely magnetic LAC
metrics~\cite{Lozanovski,Lozanovski2,LozanovskiAa}, for which the
Bianchi type is $VIII$ or $IX$, reducing to type $II$ for the $p =
\rho/5$ Collins-Stewart metric~\cite{CollinsStewart}. As $E_{\alpha
\beta}=0$ the label `purely radiative' is however not fully
appropriate for these models.

Of course the above reasoning breaks down when $n_1=n_2=0$, a case
which will be dealt with below.

\subsection{$n_2=n_3=0$}\label{sec:LRS}
Within the previous class the $n_2=n_3=0$ models are exceptional in
the sense that they are `purely electric' (by (\ref{Ricci21}) one
has $H_{\alpha\beta}=0$) and again cannot be termed `purely
radiative'. For completeness however we present the full details of
the resulting special cases.

As in (5.2) one shows that a rotation exists under which $r_2$,
$n_1$ and $\partial_3 q_3$ become zero. By (\ref{defK}) this implies
$q_3^2+K=0$, where $K$ is the same quantity as defined by (\ref{K}).
The rotation is now determined by a solution of the following
system:

\begin{equation}\label{rot3}
\eqalign{
\partial_0\varphi &= 0,\nonumber \\
\partial_1\varphi &= n_1,\nonumber \\
\partial_2\varphi &= \cos(\varphi)\sqrt{-K}-q_3,\\
\partial_3\varphi &= -\sin(\varphi)\sqrt{-K}-r_2 \nonumber
}
\end{equation}
for which again the integrability conditions are identically satisfied.

When $K<0$, one obtains
\[n_{\alpha\beta}=\left[\begin{array}{ccc} 0 &\sqrt{-K}/2&0\\
\sqrt{-K}/2&0&0\\
0&0&0\end{array}\right], \quad a_{\alpha} = \left[\begin{array}{c}0\\0\\ -\sqrt{-K}/2\end{array}\right]
\]

which is a LRS spatially homogenous Bianchi class B type $III$
model~\cite{Ellis3}.

When $K=0$ the rotation (\ref{rot3}) results in $a_{\alpha}=0$,
$n_{\alpha\beta}=0$, which gives a LRS spatially homogenous Bianchi
class A type $I$ model (equivalent to Bianchi class A type $VII_0$
\cite{Ellis3}).

A problem arises when $K>0$: $q_3$, $n_1$ and $n_2$ can still be
made zero by a rotation, but it is no longer possible to choose a
frame in which all spatial derivatives identically vanish. However
the Cartan equations can now be integrated easily and one finds the
Kantowski-Sachs perfect fluids, with metric given by
\[ds^2=-dt^2+Q(t)^2dr^2+P(t)^2[d\theta^2+\sin^2(\theta) d\phi^2]\]
and $P ( Q \ddot{P}-P \ddot{Q})+\dot{P}(Q \dot{P}-P \dot{Q})+Q=0$.
As is well known these metrics do not admit a $3$-dimensional
isometry group acting simply transitively on the
$t=\textrm{constant}$ hypersurfaces.

\section{Conclusion}
We have shown that the Bianchi class A perfect fluid models with
non-constant pressure can be uniquely characterized as geodesic
perfect fluid spacetimes which are purely radiative in the sense
that the gravitational field satisfies $\diver{H}=\diver{E}=0$,
under the assumption that also the magnetic part of the Weyl tensor
$H_{\alpha\beta}$ is diagonal in the shear-electric eigenframe
(i.e.~$[H,\, \sigma]=0$). The only possible exception arises in the
purely electric Petrov type D case of degenerate shear, where the
allowed solutions are the Kantowski-Sachs perfect fluids or the
pseudo-spherically symmetric Bianchi type III models ($K\equiv \rho
+ 3\sigma^2-\frac{1}{3}\theta^2>0$ resp.~$K<0$). The same conclusion
holds for constant pressure provided one also assumes that the fluid
is irrotational.

The next logical step is to investigate the case where $H$, $E$ and
$\sigma$ are not simultaneously diagonalizable: it is hoped that
here new classes of inhomogeneous and cosmologically interesting
solutions might turn up, or, alternatively, that one would be able
to demonstrate the remarkable result that the Bianchi class A
spacetimes are the unique (modulo the LRS exceptions above) `purely
radiative' ones. Aside from the fact that such a characterization of
the Bianchi A models would be quite neat, we feel that a further
investigation of restrictions on $\diver E$ and $\diver H$ and of
the role played by these third order quantities in general
relativistic perfect fluids might shed a new light on the issues of
PUTH and observational homogeneity.

\section*{Acknowledgment}
The authors are indebted to C.~Lozanovski for his comments on an
earlier version of the manuscript.
%Checking the spatial homogeneity
%and the Bianchi type of the LAC metrics was done with the aid of
%GRTensor II (Kayll Lake, Queen's University, Guelph, Ontario,
%Canada).

\section{Appendix}

A first set of basic equations consists of Ricci and Bianchi
equations~\cite{Ellis}, which we will use in a more formal sense and
therefore give in their covariant form (assuming
$\dot{u}_a=\omega_{a}=0$):

\subsection*{Ricci equations}

\begin{eqnarray}
 \curl\sigma_{ab}-H_{ab}=0\label{Ricci1},\\
 \dot\sigma_{<ab>}+\frac{2}{3}\theta\sigma_{ab}+\sigma_{c<a}{\sigma_{b>}}^c+E_{ab}=0\label{Ricci2},\\
 \dot\theta+\frac{1}{3}\theta^2+\sigma_{ab}\sigma^{ab}+\frac{1}{2}(\rho+3p)=0
 \label{Ricci3}.
\end{eqnarray}

In their orthonormal tetrad form (with $E_{\alpha \beta}$ and
$\sigma_{\alpha \beta}$ diagonal) the equations (\ref{Ricci1}) and
(\ref{Ricci2}) read respectively
\begin{eqnarray}
\label{orthdefE1} E_{11}=-\partial_0 \sigma_{11}+\frac{1}{3}(\sigma_{22}^2+\sigma_{33}^2-2 \sigma_{11}^2-2\theta \sigma_{11}),\\
\label{Omegas} \Omega_1 (\sigma_{22}-\sigma_{33})=0
\end{eqnarray}
and
\begin{equation}
\eqalign{
H_{11}=(\sigma_{22}-\sigma_{11}) n_3 + (\sigma_{33}-\sigma_{11})n_2, \\
\label{Ricci21} H_{12}=\frac{1}{2} \partial_3 (\sigma_{11}-\sigma_{22})+\frac{1}{2} (\sigma_{33}-\sigma_{11})r_3+\frac{1}{2}(\sigma_{33}-\sigma_{22})q_3
}
\end{equation}
and similar equations obtained by cyclic permutation of the indices.

\subsection*{Bianchi equations}
 \begin{eqnarray}\label{Bianchi1}
 \dot\rho=-(\rho+p)\theta,\\
 \label{Bianchi2}
 \diver E_a=[\sigma,H]_a+\frac{1}{3}D_a\rho,\\
 \label{Bianchi3}
 \diver H_a=-[\sigma,E]_a,\\
 \label{Bianchi4}
 \dot E_{<ab>}-\curl H_{ab}=-\theta E_{ab}+3\sigma_{c<a}{E_{b>}}^c-\frac{1}{2}(\rho+p)\sigma_{ab},\\
 \label{Bianchi5}
 \dot H_{<ab>}-\curl E_{ab}=-\theta H_{ab}+3\sigma_{c<a}{H_{b>}}^c,
 \end{eqnarray}

where the spatial dual of the commutator of tensors, $[S,T]_a$ is
defined as $[S,T]_a=\varepsilon_{abc}S^{bd}{T_d}^c$.

A second set of equations consists of those Jacobi and Einstein
equations which we use explicitly in our calculations and which we
present in their orthonormal tetrad form (using the eigenframe of $\sigma_{ab}$ and $E_{ab}$), but now with the extra simplification $\dot{u}_a=\omega_a=0$ and $\Omega_\a=0$ (see section \ref{sec:gen}). Most of these equations appear in triplets which can be obtained from each other by cyclic permutation of the
indices. For each equation we give only one representant.

\subsection*{Jacobi equations}

\begin{eqnarray}
\label{Jac234}
(j2-j4):& \quad &\partial_1(n_2+n_3)+\partial_2r_3+\partial_3q_2=\\
\nonumber &\quad & (r_1-q_1)(n_2+n_3)+r_3r_2-q_3q_2,\\
\label{Jac8910}
(j8-j10):&\quad &  \partial_0(r_1-q_1)-\partial_1(\sigma_{11}-\frac{2}{3}\theta) = -(r_1-q_1)(\sigma_{11}+\frac{1}{3}\theta),\\
\label{Jac111213}
(j11-j13):&\quad &  \partial_0(n_2+n_3)=(2 \sigma_{11}-\frac{1}{3}\theta)(n_2+n_3),\\
\label{Jac141516} (j14-j16):&\quad &
\partial_0(r_3+q_3)+\partial_3(\sigma_{11}-\sigma_{22})=-(\sigma_{33}+\frac{1}{3}\theta)(r_3+q_3).
\end{eqnarray}

\subsection*{Einstein equations}
\begin{eqnarray}\label{Ein00}
\fl(\textrm{Ein}00): &\quad & \partial_0\theta=-[(\sigma_{11}+\frac{1}{3}\theta)^2+(\sigma_{22}+\frac{1}{3}\theta)^2+(\sigma_{33}+\frac{1}{3}\theta)^2]
-\frac{1}{2}(\rho+3p),\\
\label{Ein0a}
\fl(\textrm{Ein}0\alpha): &\quad& \frac{2}{3}\partial_1\theta-\partial_1\sigma_{11}=\frac{1}{2}(r_1+q_1)(\sigma_{22}-\sigma_{33})-\frac{3}{2}(r_1-q_1)\sigma_{11},\\
\label{Einaa}
\fl(\textrm{Ein}\alpha\alpha):&\quad&-\partial_0(\sigma_{11}+\frac{1}{3}\theta)-\partial_1(r_1-q_1)+\partial_2q_2-
\partial_3r_3=\frac{1}{2}(p-\rho)\\
\nonumber & \quad &
+q_2(r_2-q_2)-r_3(r_3-q_3)-r_1^2-q_1^2+2n_2n_3+\theta(\sigma_{11}+\frac{1}{3}\theta),\\
\label{Einab}
\fl(\textrm{Ein}\alpha\beta):&\quad&-\partial_1r_2+\partial_2q_1-\partial_3(n_1-n_2)=q_1(r_2+q_2)+r_2(r_1+2q_1)\\
\nonumber & \quad & -(r_3-q_3)(n_1-n_2)+2n_3(r_3+q_3).
\end{eqnarray}

\subsection*{Commutator relations}

Finally we have the commutator relations, which can be written
succinctly as
\begin{equation}
\eqalign{ [ \partial_0 , \partial_1 ] & =
- (\sigma_{11}+\frac{1}{3}\theta) \partial_1 , \\
[ \partial_1 , \partial_2 ] & = q_2 \partial_1 +r_1
\partial_2 +(n_1+n_2) \partial_3.}
\end{equation}

\end{document}